\documentclass[prl,notitlepage, twocolumn,nofootinbib,preprintnumbers,amssymb,amsfonts,amsmath,superscriptaddress,showpacs,hyperref]{revtex4-2}

\DeclareMathAlphabet\mathbfcal{OMS}{cmsy}{b}{n}

\usepackage{graphicx}
\usepackage{dcolumn,palatino}
\usepackage{bm}
\usepackage{comment}
\usepackage{multibbl}
\usepackage{xcolor}
\usepackage{empheq}
\usepackage{amsmath}
\usepackage{amssymb}
\usepackage{ulem,xpatch}
\usepackage{float}
\usepackage{xr}
\usepackage{url}
\usepackage{soul}
\usepackage[colorlinks=true,linkcolor=blue,urlcolor=blue,citecolor=blue,pdfusetitle]{hyperref}
\usepackage[retainorgcmds]{IEEEtrantools}
\usepackage{bbm}
\def\1{\mathbbm{1}}

\newcommand{\Om}{\ensuremath{\Omega_\mathrm{m}}}
\newcommand{\Gm}{\ensuremath{\Gamma_\mathrm{m}}}
\newcommand{\Gmeas}{\ensuremath{\Gamma_\mathrm{meas}}}
\newcommand{\Gqba}{\ensuremath{\Gamma_\mathrm{qba}}}
\newcommand{\Vuc}{\ensuremath{V_\mathrm{uc}}}
\newcommand{\rfw}{\ensuremath{\mathbf{r}}}
\newcommand{\rbw}{\ensuremath{\mathbf{r}_b}}
\newcommand{\dW}{\ensuremath{d\mathbf{W}}}
\newcommand{\etad}{\ensuremath{\eta_\mathrm{det}}}
\newcommand{\bnth}{\ensuremath{\bar n_\mathrm{th}}}
\newcommand{\bncav}{\ensuremath{\bar n_\mathrm{cav}}}

\newcommand{\Vd}{\ensuremath{V_\mathrm{d}}}

\begin{document}

\title{Experimental assessment of entropy production in a continuously measured mechanical resonator}

\author{Massimiliano Rossi}
\affiliation{Niels Bohr Institute, University of Copenhagen, 2100 Copenhagen, Denmark}
\affiliation{Center for Hybrid Quantum Networks (Hy-Q), Niels Bohr Institute,University of Copenhagen, 2100 Copenhagen, Denmark}
\author{Luca Mancino}
\affiliation{Centre for Theoretical Atomic, Molecular, and Optical Physics,
School of Mathematics and Physics, Queens University, Belfast BT7 1NN, United Kingdom}
\author{Gabriel T. Landi}
\affiliation{Instituto de F\'isica, Universidade de S\~{a}o Paulo, CEP 05314-970, S\~{a}o Paulo, S\~{a}o Paulo, Brazil}
\author{Mauro Paternostro}
\affiliation{Centre for Theoretical Atomic, Molecular, and Optical Physics,
School of Mathematics and Physics, Queens University, Belfast BT7 1NN, United Kingdom}
\author{Albert Schliesser}
\affiliation{Niels Bohr Institute, University of Copenhagen, 2100 Copenhagen, Denmark}
\affiliation{Center for Hybrid Quantum Networks (Hy-Q), Niels Bohr Institute,University of Copenhagen, 2100 Copenhagen, Denmark}
\author{Alessio Belenchia}
\affiliation{Centre for Theoretical Atomic, Molecular, and Optical Physics,
School of Mathematics and Physics, Queens University, Belfast BT7 1NN, United Kingdom}

\date{\today}

\begin{abstract}
The information on a quantum process acquired through measurements plays a crucial role in the determination of its non-equilibrium thermodynamic properties. We report on the experimental inference of the stochastic entropy production rate for a continuously monitored mesoscopic quantum system. We consider an optomechanical system subjected to continuous displacement Gaussian measurements and characterise the entropy production rate of the individual trajectories followed by the system in its stochastic dynamics, employing a phase-space description in terms of the Wigner  entropy. Owing to the specific regime of our experiment, we are able to single out the informational contribution to the entropy production arising from conditioning the state on the measurement outcomes. Our experiment embodies a significant step towards the demonstration of full-scale control of fundamental thermodynamic processes at the mesoscopic quantum scale.
\end{abstract}

\maketitle

The fundamental connections between information and thermodynamics dates back to the seminal contributions by Maxwell, Szilard, and Landauer~\cite{Parrondo2015}.
The process of acquiring information can impact the entropic balance of a given physical process. Such information must thus be accounted for when formulating the second-law, and considered on equal footing to other thermodynamic quantities, such as heat and work.
This is particularly relevant for processes involving microscopic systems, which are fundamentally dominated by fluctuations: the acquisition of information through measurements introduces additional stochasticity and makes the overall process strongly dependent on the monitoring methodology. 

Let us briefly illustrate the building blocks of the formulation of thermodynamics at the stochastic level. The changes $dS$ in the entropy of a system subjected to a process can be attributed both to a  flow of entropy $\phi$ between the system and its surroundings and to a contribution $\pi$ associated to the irreversible production of entropy~\cite{RevModPhys.20.51}. We can thus write
\begin{equation}\label{dS}
    dS = \phi + \pi,
\end{equation}
where both $\phi$ and $\pi$ are stochastic quantities that fluctuate with each repetition of the experiment. Averaging over many realisations yields the entropy flux and production rates, $\Phi$ and $\Pi$ respectively. 
The second law enforces $\Pi \geqslant 0$ with $\Pi = 0$ when the system is in equilibrium. The introduction of measurement and feedback processes profoundly affects the above statements, as it has been studied in both classical~\cite{Sagawa2010,Sagawa2012,Sagawa2013,Ito2013,Horowitz2014b} and quantum contexts~\cite{Sagawa2008,*Sagawa2009,Ito2013,Funo2013,benoist2018}.
Experimental assessments of such modifications have been reported recently in a variety of systems, including classical Brownian particles~\cite{toyabe2010experimental}, superconducting qubits~\cite{Masuyama2017,Cottet2017}, trapped ions~\cite{Xiong2018,Yan2018} and nuclear magnetic resonance~\cite{Camati2016,Peterson2016}.
Frameworks for describing the dynamics of {\it continuously monitored} quantum systems have been developed ~\cite{breuer2002, wisemanQuantumMeasurementControl2010, PhysRevLett.116.080403,dressels2017,Elouard2017a} and fundamental fluctuation theorems involving heat, work, and entropy for continuously monitored quantum two-level systems have been studied~\cite{Stefano2018,PhysRevLett.121.030604,PhysRevLett.123.020502,naghiloo2017heat}, including assessments at the single-trajectory level~\cite{murch2013observing,weber2014mapping}.

When assessing a monitored system, one must distinguish between the unconditional evolution and the  dynamics conditioned on the measurement records~\cite{Ito2013,Funo2013,Strasberg2019a}. It can be shown that the average entropy flux rate of the conditional dynamics $\Phi_c$ equals the unconditional one $\Phi_\text{uc}$~\cite{belenchia2019entropy}. However, the same is not true for the entropy production rate that governs the irreversibility of the process. Acquiring information can only make the process more reversible, so that the average entropy production {$\int^t \Pi_c\,d\tau$} of the {\it conditional trajectories} will be smaller than that of the unconditional one $\int^t \Pi_{\text{uc}}\,d\tau$. 
Their difference is precisely associated with an information-theoretic term and can be written as~\cite{belenchia2019entropy,Horowitz2014b,Strasberg2019a} 
\begin{equation}\label{2_pis}
    \Pi_c = \Pi_\text{uc} + \dot{\mathcal{I}},
\end{equation}
where $\dot{\mathcal{I}}$ is the net rate at which information is acquired through measurement. This term thus captures the fundamental effect of the measurement, providing a valuable link between information and thermodynamics.

Despite its importance, thus far there has been no experimental assessment of the influence of such information theoretic contribution, arising from quantum measurements, to the entropy production.
In this paper, we fill such gap by reporting the  experimental observations of the impact of weak continuous measurements on the non-equilibrium thermodynamics of a mesoscopic mechanical resonator~\cite{PhysRevLett.123.163601}. 
Our system consists of a nanomechanical resonator coupled to an optical cavity and exposed to the effects of both electromagnetic and phononic environments. We continuously monitor its position by means of homodyne measurements on the output optical field~\cite{PhysRevLett.123.163601}. 
Combining the phase-space formalism laid down in Ref.~\cite{belenchia2019entropy}  with state retrodiction methods~\cite{PhysRevLett.123.163601}, we are able to characterise the entropy production at the level of {individual} quantum trajectories. It should be noted that, in the phase-space formalism, we chose the Wigner entropy  as the entropic measure which, despite its limitations to Gaussian dynamics, presents several advantages when considering non-standard thermodynamics setups~\cite{PhysRevLett.118.220601} (see the discussion in the following).
Remarkably, we are able to single-out precisely the contribution of the measurement influence to the entropy production.

Our experiment probes both the relaxation dynamics and the steady-state. 
The latter, in particular, configures an {\it informational steady-state}, where information acquired from the measurement is constantly counterbalancing noise introduced by the environment. 
In addition to the net rate of information gain $\dot{\mathcal{I}}$,  we are also able to single-out the {\it differential} information gain $\mathcal{G}(t)$, which represents the rate at which information must be acquired in each small time step in order to maintain this steady-state. 
Our work thus embodies a step forward towards the full characterisation of quantum mesoscopic irreversibility and its control via suitably arranged measurements.

\begin{figure}[t!]
    \centering
    \includegraphics[width=\columnwidth]{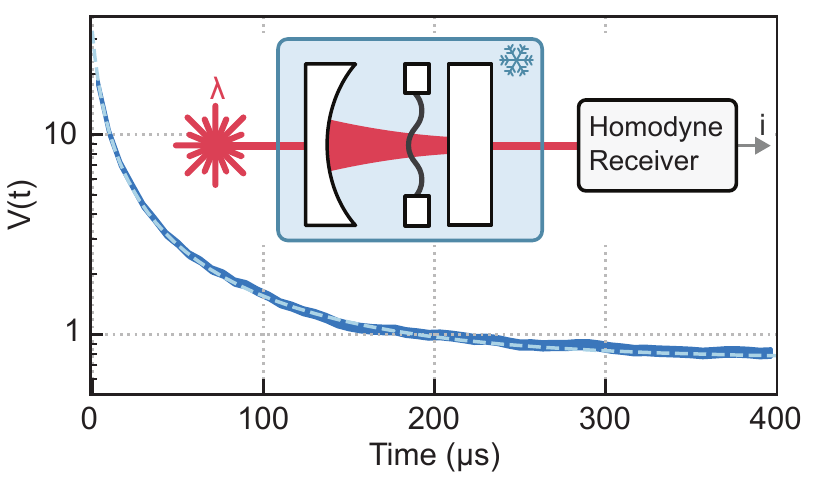}
    \caption{{\bf Conditional mechanical evolution.}
    Measured conditional variance $V(t)$ (blue line), from the initial unconditional value $\Vuc \approx 34$ to the steady-state $V_\text{ss}\approx0.8$ . The dashed line is a theoretical prediction. The inset shows a sketch of the experimental system, which comprises a cryogenic optomechanical cavity resonantly driven by a coherent probe laser. The mechanical resonator is in thermal contact with two baths: a thermal, cryogenic bath and the optical bath. The output field is continuously monitored by means of a homodyne receiver. The photocurrent $i$ is used to estimate the conditional mechanical state. } 
    \label{fig:unc_cond_state}
\end{figure}

\noindent
\textbf{Experimental set-up.} The experimental system is provided by an ultracoherent soft-clamped membrane resonator [cf. inset of Fig.~\ref{fig:unc_cond_state}]. The central defect, embedded in a phononic crystal, supports a localised, ``soft-clamped'' mechanical mode~\cite{RossiNature} at the resonance frequency $\Omega_m/(2\pi)= 1.14$~MHz. Once cooled to a temperature of $T=11$~K, we find for this mode a quality factor $Q=\Om/\Gm=1.03\times10^9$, where $\Gm$ is the energy dissipation rate. The mechanical system is dispersively coupled to the frequency of a Fabry-Perot cavity mode (linewidth $\kappa/(2\pi) = 18.5$~MHz), with vacuum optomechanical coupling rate $g_0/(2\pi)=129$~Hz. The cavity mode is pumped by an external probe laser to an averaged photon occupancy $\bncav$. We assume that the semi-classical steady-state of the non-linear dynamics has been reached and, when speaking of different modes, we refer to the fluctuations around such a mean steady-state, as it is common practice \cite{Aspelmeyer2014}. In this linearised interaction regime, the effective, multi-photon optomechanical coupling, enhanced by the average cavity photon occupancy, is $g=g_0\sqrt{\bncav}$ and the fluctuations of the  system evolve according to a Hamiltonian that is quadratic in the system's fluctuations. This ensures that all the states remain Gaussian.

We use an auxiliary light field to stabilise the system and provide pre-cooling of other mechanical modes. Such beam also introduces additional damping and cooling on the mode of interest, effectively changing its thermal environment. In addition, any small detuning of the probe beam from the cavity resonance causes additional damping. We account for these effects by introducing the effective energy damping rate and bath occupancy $\Gm/(2\pi)=19$~Hz and $\bnth=14$, respectively. The total thermal decoherence rate is thus $\gamma=\Gm(\bnth+1/2)=2\pi\times265$~Hz.

The quantum measurement is performed by imprinting, through the optomechanical interaction, the mechanical displacement in the phase quadrature of the probe laser.
Such quadrature is measured by a phase-sensitive measurement of the output field, implemented using a balanced homodyne receiver with detection efficiency $\eta_\text{det}=74\%$ [cf. inset of Fig.~\ref{fig:unc_cond_state}].

Our experiment operates in the non-resolved-sideband regime $\Omega_m\ll\kappa$, which enforces a separation of time-scales and allows the cavity mode to be adiabatically eliminated~\cite{PhysRevLett.123.163601}. 
In a frame rotating at frequency $\Omega_m$ and within the rotating wave approximation, the conditional dynamics of the mechanical mode alone is well described by the stochastic master equation (SME)~\cite{PhysRevLett.107.213603,doherty2012quantum} \begin{equation}
    \label{adi}
d\rho_c=\left({\cal L_\text{th}}+{\cal L_\text{qba}}+{\cal L_\text{stoc}}\right)\rho_c dt.
\end{equation}
The term $\mathcal{L}_{\rm th}\rho=\Gamma_m (\bnth+1)\mathcal{D}[\hat c]\rho+\Gamma_m \bnth\mathcal{D}[\hat c^\dag]\rho$ in Eq.~\eqref{adi} describes the contact with the effective mechanical phonon bath.  
Here $\hat{c}=(\hat{X}+i\hat{Y})/\sqrt{2}$ is the annihilation operator of the mechanical mode (written in terms of  quadratures $\hat{X}, \hat{Y}$) and $\mathcal{D}[\hat A]\rho= \hat A\rho \hat A^\dag-\{\hat A^\dag \hat A,\rho\}/2$. 
The quantum measurement backaction is described by ${\cal L_\text{qba}}\rho_c=\Gamma_{\rm{qba}} \left(\mathcal{D}(\hat c)+\mathcal{D}(\hat c^\dag)\right)\rho_c$, which results in radiation pressure force fluctuations with a decoherence rate
$\Gqba= 4g^2/\kappa\approx 2\pi\times 0.36$~kHz. 
Finally, the term
\begin{equation}
{\cal L_\text{stoc}}\rho_c=   
\sqrt{\eta_{\rm{det}}\Gamma_{\rm{qba}}}\left(\mathbfcal{H}\rho_c\right)\cdot d\text{\bf W}
\end{equation}
with the vectors $\mathbfcal{H}\rho_c=(\mathcal{H}(\hat X)\rho_c\, ,\mathcal{H}(\hat{Y})\rho_c)^T$ and $d\text{\bf W}=(dW_X\, ,dW_Y)^T$, describes the stochastic contribution to the dynamics stemming from conditioning upon the measurement outcomes, with $\mathcal{H}(\hat A)\rho=\hat A\rho+\rho \hat A^\dag-\langle \hat A+\hat A^\dag\rangle_\rho \rho$, and $dW_{X,Y}$ independent real Wiener increments \cite{wisemanQuantumMeasurementControl2010, doherty2012quantum}.

The unconditional mechanical dynamics is retrieved by neglecting ${\cal L_\text{stoc}}\rho_c$. The corresponding Gaussian steady-state density matrix is characterised by vanishing first cumulants, $\mathbf{r}_\mathrm{uc} = (\langle \hat{X} \rangle_\text{uc}\, ,\langle \hat{Y}\rangle_\text{uc})^T = \mathbf{0}$, and a diagonal covariance matrix with elements $\Vuc = \langle \hat{X}^2 \rangle_\text{uc} = \langle \hat{Y}^2 \rangle_\text{uc} =\bnth + 1/2 + \Gqba/\Gm$. We use such state density matrix as the initial preparation in all experiments reported below. This is the natural steady-state of the optomechanical system, thus its preparation requires only to wait for the initial brief transient to decay, before conditioning upon the measurement outcomes.
The conditional dynamics described by Eq.~\eqref{adi}, on the other hand, has both first and second cumulants evolving non-trivially according to 
\begin{IEEEeqnarray}{rCl}
\label{eq:first_moment}
d\rfw(t) &=& - \frac{\Gamma_m}{2}\rfw dt + \sqrt{4 \eta_{\rm{det}} \Gamma_\text{qba}} V(t) d\mathbf{W}, 
\\
\label{eq:second_moment}
\dot{V}(t) &=& \Gamma_m (V_\text{uc} - V(t))
- 4 \eta_{\rm{det}} \Gamma_{qba} V(t)^2.
\IEEEeqnarraynumspace
\end{IEEEeqnarray}
The ensuing dynamics gives the covariance ${\bm V}(t)=V(t)\openone$ where we have introduced the  identity matrix $\openone$ and the $c$-number variance $V(t) = \langle \hat{X}^2 \rangle  - \langle \hat{X} \rangle^2 = \langle \hat{Y}^2 \rangle - \langle \hat{Y}\rangle^2$~\cite{SM}. 
The first cumulants thus evolve stochastically, while the second ones obey a deterministic non-linear evolution. It should be noted that, the process entailed by this model is dynamically stable, as it can be easily verified following the criteria discussed in~\cite{PhysRevLett.94.070405}. This ensures the convergence of any quantity integrated over long-time windows. 

The last term in Eq.~(\ref{eq:second_moment}) is associated with the information acquired by the measurement, and we dub it {\it innovation}. It is non-positive as  acquired information can never increase the uncertainty about the mechanical motion. 
According to Eq.~(\ref{eq:second_moment}), the initial unconditional variance $V_\text{uc}$ evolves into the conditional steady-state  value 
$V_{ss} = - \mu + \sqrt{\mu(\mu+ 2 V_\text{uc})}$ with $\mu = \Gamma_m /(8 \eta_{\rm{det}} \Gamma_\text{qba})$.
Owing to the innovation term, $V_{ss} \leqslant V_\text{uc}$ given that $\mu > 0$.
The continuous weak measurements thus lead to a conditional steady-state density matrix with a higher purity than the unconditional one \cite{maassen2006, barchielli2003}.
This is an instance of measurement-based cooling and was experimentally demonstrated in~\cite{PhysRevLett.123.163601}.
The conditional first cumulants $\rfw(t)$ in Eq.~(\ref{eq:first_moment}) are related to the experimental homodyning measurement outcomes $\mathbf{i}(t)$ through $\mathbf{i}(t) dt=\sqrt{4\eta_{\rm{det}}\Gamma_{\rm{qba}}}\rfw(t) dt+d\mathbf{W}$. 
In practice, the latter relation is used to express $d\mathbf{W}$ in terms of the outcomes, $\mathbf{i}(t)$, and substituted in Eq.~(\ref{eq:first_moment}). This yields a recursive relation used to experimentally filter the data $\mathbf{i}(t)$ to obtain the first cumulant $\mathbf{r}(t)$.
The conditional variance $V(t)$, however, evolves independently of the specific measurement outcomes. 
To assesses it experimentally we thus employ a prediction-retrodiction method~\cite{PhysRevLett.123.163601}, which reconstructs $V(t)$ by combining data on $\rfw(t')$ acquired at earlier ($t' < t$) {\it and} {later} times ($t'>t$). Such future outcomes can be used to obtain a retrodicted trajectory, $\rbw(t)$, ~\cite{Zhang2017} using an experimental filter similar to what has been derived fro $\mathbf{r}(t)$ ~\cite{SM}. 
The fluctuations of the difference $\textbf{d}(t) = \rfw(t)-\rbw(t)$ over an ensemble of independent realisations can be shown to be directly connected to $V(t)$ according to the relation~\cite{SM}  
\begin{equation}
\label{variance}
V_d(t) = V(t) + V_\text{ss} + \Gm/(4\eta_{\rm{det}} \Gamma_\text{qba}).
\end{equation}
In the limit of high-cooperativity ($\Gqba\gg\Gm$) and large detection efficiency ($\eta_{\rm{det}}\approx1$) the last term can be neglected.
The value of $V_\text{ss}$ and $V(t)$ are then readily obtained as $V_\text{ss}=V_d(\infty)/2$~\cite{PhysRevLett.123.163601} and $V(t)=V_d(t)-V_\text{ss}$, respectively. 

Figure~\ref{fig:unc_cond_state} shows the evolution of $V(t)$ from the initial unconditional value $V_\text{uc}$, all the way to the steady-state value $V_\text{ss}$. The experimental data compare very well to the theoretical prediction provided by Eq.~(\ref{eq:second_moment}), thus strongly corroborating the suitability of our model. 

\noindent
\textbf{Entropy production along individual trajectories.} We are now in a position to assess the thermodynamics of the system at the level of individual quantum trajectories. 
Our setup is not a standard thermodynamic system due to the presence of the optical cavity, which acts as a non-thermal bath. 
The usual formulation of entropy production thus does not apply. 
Despite this, it is possible to employ an alternative put forth in Ref.~\cite{PhysRevLett.118.220601}, which makes use of quantum phase-space methods and is adequate for the description of Gaussian dynamics.
This approach has already been successfully applied to the experimental characterisation of the mean entropy production in the dynamics of open mesoscopic systems~\cite{PhysRevLett.121.160604}. 
In Ref.~\cite{belenchia2019entropy}, the method was extended to account for the presence of quantum-limited detectors continuously monitoring the system.

\begin{figure}
    \centering
    \includegraphics[width=0.95\columnwidth]{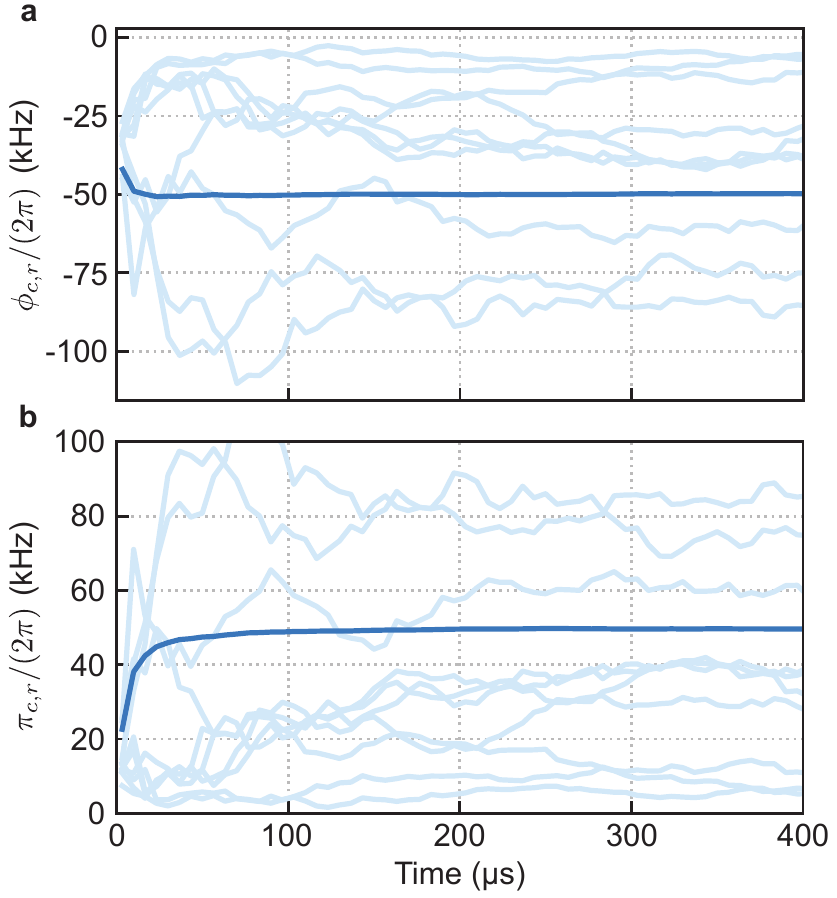}
    \caption{{\bf Stochastic entropy flux and production rates. }
    {\bf a,} The stochastic entropy flux rates (light blue) for a sample of 10 trajectories. The dark blue line is the ensemble average over all the trajectories.
    {\bf b,} The stochastic entropy production rates (light blue) and the ensemble average (dark blue), for the same sample of trajectories.}
    \label{fig:stoc_entropy}
\end{figure}

When applied to our experimental endeavours~\cite{SM}, such theoretical framework shows that the conditional entropy flux and production rates, defined in Eq.~(\ref{dS}), can be written in terms of the first and second cumulants as
\begin{IEEEeqnarray}{rCl}
\nonumber
\phi_{c, \rfw}&=&\frac{\Gamma_m}{n_\text{th}+1/2}\left[(n_\text{th}+1/2)-\theta(t)\right] - 4 \Gamma_\text{qba} \theta(t),\\
&&\label{stochastic}\\
\pi_{c, \rfw}&=&\Gamma_m\left[\frac{\theta(t)}{n_\text{th}+1/2}+\frac{V_{\rm{uc}}}{V(t)}-2\right] + 4 \Gamma_\text{qba}\left[ \theta(t) {-} \eta_{\rm{det}} V(t)\right],
\nonumber
\end{IEEEeqnarray}
where $\theta(t) = V(t)+\bm{r}(t)^\text{T} \bm{r}(t)/2$ encompasses all the stochastic contributions [cf.~Eq.~(\ref{eq:first_moment})].
We can experimentally reconstruct such quantities by means of the measured stochastic trajectories $\rfw(t)$ and the inferred conditional variance $V(t)$. We show in Fig.~\ref{fig:stoc_entropy} some realisations of the stochastic entropy flux and production rates. Despite the low thermal occupancy of $n_{\rm{th}}\approx 14$ phonons, these quantities fluctuate substantially, highlighting the essential role of fluctuations in the thermodynamics of the system. 

We also average them over 3600 trajectories, yielding the conditional flux and production rates $\Phi_c = \mathbb{E}(\phi_{c,\bm{r}})$ and $\Pi_c = \mathbb{E} (\pi_{c,\bm{r}})$, which are shown in Fig.~\ref{fig:stoc_entropy}, dark blue.
These quantities can be readily computed from our model by noting that, owing to Eqs.~(\ref{eq:first_moment}) and (\ref{eq:second_moment}) and  our choice of initial conditions, we have  
$\mathbb{E}\big[\theta(t)\big] = V_\text{uc}$. 
From Fig.~\ref{fig:stoc_entropy} we gather that both $\Phi_c$ and $\Pi_c$ relax monotonically towards the new steady-state values. However, even at the steady-state, the entropy production rate $\Pi_c$ does not vanish due to the non-equilibrium nature of the stationary state, where the effects of the thermal bath, measurement backaction, and information gain compete with each other. 

\begin{figure}
    \centering
    \includegraphics[width=0.9\columnwidth]{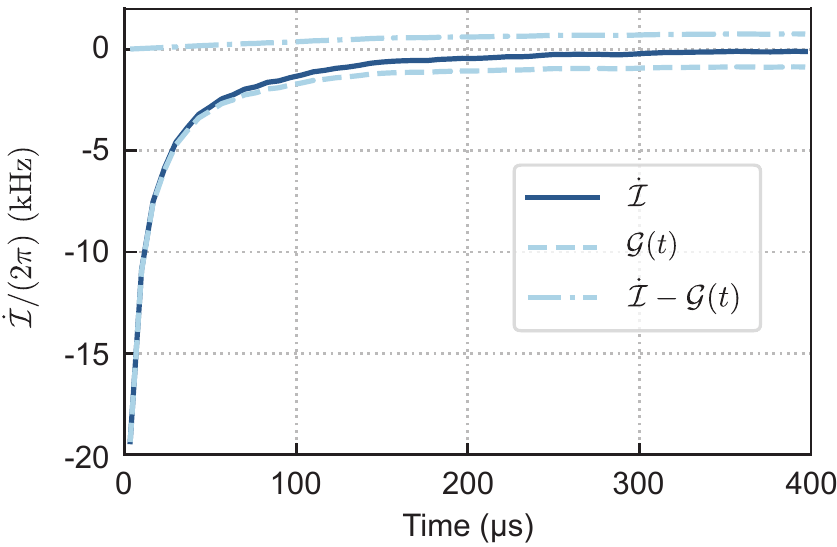}
    \caption{{\bf Informational contribution to the entropy production rate.}
    We obtain the informational contribution (dark blue) from the entropy production. The dashed (dot-dashed) line is the differential gain of information due to the measurement (loss of information due to noise input by the phonon bath).}
    \label{fig:informational_term}
\end{figure}

\noindent
\textbf{Information gain.}  
The influence that monitoring the system has on the irreversibility of the dynamics is encoded in the mismatch between the conditional entropy production rate $\Pi_c$ and the unconditional one $\Pi_\text{uc}$ [cf.  Eq.~(\ref{2_pis}]. Such mismatch is quantified by  the net rate of information gain achieved through measuring 
\begin{equation}
    \label{Idot}
 \dot{\mathcal{I}}=\Gm \left({\Vuc}/{V(t)}-1 \right)-4\etad \Gamma_\text{qba} V(t).
\end{equation}
The temporal behaviour of $\dot{\mathcal{I}}$ reconstructed from the experimental data is shown in Fig.~\ref{fig:informational_term}. As in our  case the system is prepared in the steady-state of the unconditional dynamics, the first and second cumulants in the absence of monitoring remain constant in time, and the unconditional rate of entropy production keeps the value $\Pi_\text{uc} = \Gamma_m\big[ V_\text{uc}/(n_\text{th}+1/2) -1\big] + 4 \Gamma_\text{qba} V_\text{uc}$ (cf.~\cite{SM} for further details). We can thus subtract such value from $\Pi_{\rm c}$ in Fig.~2 to obtain the net rate of acquired information due to the continuous monitoring.

As the quantity $-\int^\infty_0\dot{\mathcal{I}}dt$ quantifies the mutual information between system and  detector~\cite{belenchia2019entropy}, and given that $\dot{\mathcal{I}}$ vanishes in the (conditional) steady-state [cf.~Fig.~\ref{fig:informational_term}], such quantity tends to a constant in the long-time limit. 
This is intuitively understood from the fact that, in the steady-state, monitoring the system does not add any additional information. 
If, however, the monitoring process suddenly stops, the conditional steady-state will not be sustained and the system will heat-up back towards $V_\text{uc}$. 
Constant monitoring is thus necessary to maintain the conditional steady-state. 
In other words, even at the steady-state, information is constantly being acquired, but noise is constantly being introduced by the phonon bath. 
It is thus interesting to identify which of the terms in $\dot{\cal I}$ is responsible for the incremental gains of information required to maintain the conditional steady-state. 

This concept can be readily understood from inspecting  Eq.~\eqref{Idot}, which consists of the competition between the noise introduced by the phonon bath (at rate $\Gamma_\text{m}$) and the gain of information (proportional to the detection efficiency $\eta_{\rm{det}}$).
We can thus quite naturally introduce the {\it differential gain} $\mathcal{G}(t) := - 4 \eta_{\rm{det}} \Gamma_\text{qba} V(t)$ and notice that, in light of the interpretation of the last term in Eq.~(\ref{eq:second_moment}) as an innovation rate, $\mathcal{G}(t)$ is the contribution of this innovation to $\dot{\mathcal{I}}$.
The behaviour of $\dot{\cal I}$ and ${\cal G}(t)$ inferred from the experimental data are shown in Fig.~\ref{fig:informational_term}:
the initial closeness of $\mathcal{G}(t)$ to $\dot{\mathcal{I}}$ suggests that the early stages of the dynamics are strongly affected by the differential information gain. As the dynamics approaches the steady-state, however, the contribution from $\mathcal{G}(t)$ become less significant. However,  while $\dot{\mathcal{I}}\to0$, $\mathcal{G}(t)$ tends to the (in general small) non-null value $\mathcal{G}(\infty) = - 4 \eta_{\rm{det}} \Gamma_\text{qba} V_\text{ss}$, which thus represents the gain of information per unit time that the detector must acquire in order to maintain the steady-state.

\noindent
{\bf Conclusions.}
We have investigated the effects of weak continuous measurements on the thermodynamics of a mesoscopic mechanical system. By employing a phase-space formalism~\cite{belenchia2019entropy} and the retrodictive techniques used in Ref.~\cite{PhysRevLett.123.163601}, we have connected pivotal thermodynamic quantities, such as entropy production and flux rates along individual dynamical trajectories, to accessible experimental data. The working point of our experiment has enabled us to single out the contributions to the entropy production of the system due solely to the information acquired by monitoring the system. Such contribution decreases in time as the system reaches a non-equilibrium steady state.

Our endeavours demonstrate the key role played by measurements in influencing the energetics of a quantum system. Remarkably, they  showcase the intricate interplay between fundamental energy-exchange processes and information in setting up (and sustain) the dynamical and steady-state features of a process. Such influences can be further explored by assessing whether the control of informational terms to entropy production stemming from suitable measurement strategy could be used as an effective tool for quantum state engineering~\cite{cardona2020}. Another interesting direction would address composite systems endowed with initial quantum correlations and the experimental study of their effects, in conjunction with continuous monitoring, on the thermodynamics of the systems.

\noindent
{\bf Acknowledgements.} AB acknowledges  hospitality at the Institute for Theoretical Physics and the ``Non equilibrium  quantum  dynamics'' group, Universit\"{a}t Stuttgart, where part of this work was carried out. We acknowledge financial support from the European Research Council project Q-CEOM (grant nr. 638765), Danish National Research Foundation (Center of Excellence “Hy-Q”), the EU H2020 FET proactive project HOT (grant nr. 732894), Fondazione Angelo Della Riccia, the S\~ao Paulo Research Foundation FAPESP (grants 2017/50304-7, 2017/07973-5 and 2018/12813-0), the Brazilian CNPq (grant INCT-IQ 246569/2014-0), the MSCA IF project pERFEcTO (grant nr. 795782), the H2020-FETOPEN-2018-2020 project TEQ (grant nr.~766900), the DfE-SFI Investigator Programme (grant 15/IA/2864), COST Action CA15220, the Royal Society Wolfson Research Fellowship (RSWF\textbackslash R3\textbackslash183013), the Royal Society International Exchanges Programme (IEC\textbackslash R2\textbackslash192220), the Leverhulme Trust Research Project Grant (grant nr.~RGP-2018-266), the UK EPSRC.

%

\newpage
\clearpage
\onecolumngrid
\setcounter{equation}{0}
\setcounter{figure}{0}
\setcounter{table}{0}
\setcounter{page}{1}
\makeatletter
\renewcommand{\theequation}{S\arabic{equation}}
\renewcommand{\thefigure}{S\arabic{figure}}
\renewcommand{\bibnumfmt}[1]{[S#1]}

\begin{center}
  {\bf{\Large  Supplemental Materials:}\\{\large Experimental assessment of entropy production \\in a continuously measured mechanical resonator}}
  \vspace{5mm}\\  
  {\normalsize Massimiliano Rossi,$^{1,2}$ Luca Mancino,$^{3}$ Gabriel T. Landi,$^{4}$,\\\vspace{1mm}
    Mauro Paternostro,$^{3}$ Albert Schliesser,$^{1,2}$ and Alessio Belenchia$^{3}$}\\\vspace{3mm}
{\it $^1$  Niels Bohr Institute, University of Copenhagen, 2100 Copenhagen, Denmark\\\vspace{0.5mm}
  $^2$  Center for Hybrid Quantum Networks (Hy-Q), Niels Bohr Institute,\\ University of Copenhagen, 2100 Copenhagen, Denmark\\\vspace{0.5mm}
  $^3$ Centre for Theoretical Atomic, Molecular, and Optical Physics, \\ School of Mathematics and Physics, Queens University, Belfast BT7 1NN, United Kngdom\\\vspace{0.5mm}
  $^4$ Instituto de F\'isica, Universidade de S\~{a}o Paulo, CEP 05314-970, S\~{a}o Paulo, S\~{a}o Paulo, Brazil}
\end{center}

In this supplementary material we include a discussion on the theoretical derivation of the entropy production rate formulas used in the main text. We also include the experimental parameters used in the experiment, as well as details of the retrodiction techniques employed to reconstruct the dynamical evolution of the covariance matrix of the system from the experimental data. 

\section{System parameters}

We give here a brief description of the main experimental parameters and how they are measured. A summary is provided in Tab.~\ref{t:parameters}.

\begin{figure}[h]
    \centering
    \includegraphics{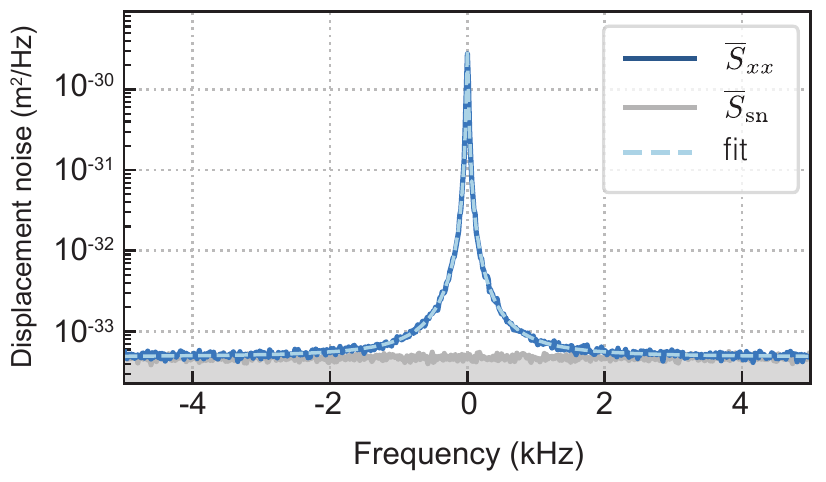}
    \caption{\textbf{Calibrated displacement spectrum.} Homodyne photocurrent power spectral density, calibrated in displacement units (blue), and a Lorentzian fit (dashed light blue). The shot noise PSD (gray) is acquired when the light from the optomechanical cavity is blocked. The offset in the horizontal axis is the mechanical resonance frequency $\Omega_m/(2\pi)=1.140$~MHz.}
    \label{fig:SM_Sxx}
\end{figure}

We estimate the power spectral density (PSD) from the homodyne photocurrent $i(t)$. The fluctuating mechanical motion appears as a Lorentizian peak around the resonance frequency $\Omega_m/(2\pi)=1.140$~MHz, as shown in Fig.~\ref{fig:SM_Sxx}.
This PSD is calibrated into displacement units via a phase-modulation technique~\cite{Gorodetsky2010_SI}, combined with the independent measured vacuum optomechanical coupling $g_0/(2\pi)=129$~Hz~\cite{PhysRevLett.123.163601_SI}. From a Lorentzian fit  we get the mechanical linewidth $\Gm/(2\pi)=19.0$~Hz and the total, unconditional occupation $\overline{n}_\text{uc}=\Vuc-1/2=33$.

We infer the multi-photon optomechanical coupling $g/(2\pi) = 40.8$~kHz from the measurement of the output optical power used in the experiment. The quantum backaction rate is $\Gqba = 4 g^2/\kappa = 2\pi\times 360$~Hz. The total detection efficiency, including cavity overcoupling, optical losses from the cavity output to the homodyne detector, photodiodes quantum efficiencies and homodyne interference visibility, is $\eta_\text{det}=74\%$. We estimate a measurement rate of $\Gmeas = \eta_\text{det} \Gqba = 2\pi\times 266$~Hz.

From the unconditional variance $\Vuc$ and the classical cooperativity $\mathcal{C}=\Gqba/\Gm$ we estimate the average number of thermal phonons $\bnth = \Vuc - \mathcal{C}-1/2=14$, consistent with a bath temperature of $T=11$~K when the dynamical backaction from the probe and auxiliary lasers are taken into account.

\begin{table}[H]
\centering
  \begin{tabular}{llll}
    \hline
    Symbol & Definition & Name &  Value\\
    \hline
  $\Gm$ & & Effective mechanical damping rate &  $2\pi\times19.0$~Hz  \\
  $\bnth$ & & Effective thermal bath occupation &  14 \\
  $\Gmeas$ & $\eta_\text{det}\Gqba$ & Measurement rate & $2\pi\times266$~Hz \\
  $\mathcal{C}$ & $\Gqba/\Gm$ & Classical cooperativity & 19\\
  $\Vuc$ & $\bnth + \mathcal{C} + \frac{1}{2}$& Unconditional variance  & 33.5\\
\end{tabular}
\caption{{\bf Summary of the main experimental parameters.}}\label{t:parameters}
\end{table}

\section{\label{sec:level1} Irreversible entropy production rate for linear quantum systems}

Consider a general Gaussian quantum system subject to continuous measurements. The general form of the master equation describing the Markovian conditioned dynamics of the system is given by Eq.[3] of Ref.~\cite{PhysRevLett.94.070405_SI}. It should be noted that, the master equation of the system studied in the main text is exactly of this form. However, for the case in which we are interested it suffices to consider the less general master equation (see also Ref.~\cite{jacobs2006straightforward_SI})  
\begin{equation}\label{SMEgen}
    d\rho_c=-i[H,\rho_c]dt+\sum_{\ell=1}^L\mathcal{D}[c_\ell]\rho_c dt+\sum_{\ell=1}^L\sqrt{\eta_\ell}\mathcal{H}[c_\ell]\rho_c dW_\ell,
\end{equation}
where $\rho_{c}$ is the conditional density matrix, $\hat{c}_\ell$ are arbitrary bounded operators, $\mathcal{D}[c]=c\rho c^\dag-\{c^\dag c,\rho\}/2$, and $\mathcal{H}[c]\rho=c\rho+\rho c^\dag-\rm{Tr}[\rho (c+c^\dag)]\rho$. Here $L$ is the number of output channels which can be measured with efficiencies $\eta_\ell$, and $dW_\ell$ are real, independent, Wiener increments. 

This master equation corresponds to a stochastic Fokker-Planck (sFP) equation in phase-space for the dynamics of the Wigner function, $W$, associated to the state density matrix of the system. Given the Gaussian nature of the problem at hand, the dynamics of the system is also fully characterised by the equations for the first two cumulants of the Wigner distribution,
\begin{eqnarray}
d\rfw(t) &=& \sum_{\ell=1}^L A_\ell\rfw(t) dt +\sum_\ell(V(t)C_\ell^T+\Gamma_\ell^T)d\mathbf{W},\label{r_riccatti}\\
\dot{V}(t) &=&\sum\limits_{\ell=1}^L \left(A_{\ell}V(t)+V(t)A_\ell+D_\ell\right) -\sum_\ell\chi_\ell(V(t)).\label{V_riccatti}
\end{eqnarray}
Here $\rfw(t)$ are the first cumulants, $C$ and $\Gamma$ are matrices describing the measurement process, $d\mathbf{W}$ is a vector of independent Wiener increments~\cite{PhysRevLett.94.070405_SI}, $V(t)$ is the quadrature covariance matrix (CM) and the term $\chi(V(t))$, called the innovation matrix, is a quadratic function of $V$ quantifying the rate at which information about the system is learned through the measurement. The sum in Eqs.~(\ref{r_riccatti})-(\ref{V_riccatti}) refers to the different environments (output channels) acting on the system; splitting the contributions of $A$, $D$ and $\chi$ in this way will be convenient in what follows.

Following Ref.~\cite{belenchia2019entropy_SI}, and using as entropic measure the Wigner entropy~\cite{PhysRevLett.118.220601_SI}, the thermodynamics of the system can be characterised by establishing the corresponding  entropy production and flux rates. Let us first focus on the unconditional evolution, i.e., when the measurement outcomes are not recorded (equivalently, when the detectors' efficiency vanishes). 
This is obtained from~\eqref{SMEgen} by setting $\eta_\ell = 0$. 
In this case, the entropy production and flux rates are deterministic quantities, given by~\cite{PhysRevLett.118.220601_SI} 
\begin{equation}
\Phi_{\rm{uc}} =-2\int d^{2n}\mathbf{x} \sum_\ell J^{T}_{\rm{irr},\ell}D_{\ell}^{-1}A_{\rm{irr},\ell}\mathbf{x},\quad \Pi_{\rm{uc}}=2\int \frac{d^{2n}\mathbf{x}}{W_{\rm{uc}}}\sum_\ell J_{\rm{irr},\ell}^TD^{-1}_\ell J_{\rm{irr},\ell},
\end{equation}
where $A_{\rm{irr},\ell}$ is the irreversible part of the drift matrix stemming from the corresponding dissipator $\mathcal{D}[c_\ell]$, and  $J_{\rm{irr},\ell}=A_{\rm{irr},\ell}\mathbf{x} W-({D_\ell}/{2})\nabla W$. More explicitly,
\begin{align}\label{ucthermo}
     & \Phi_{\rm{uc}}=-{\rm{Tr}}[A_{\rm{irr}}]-\sum_\ell\left(2{\rm{Tr}}[A_{\rm{irr},\ell}^TD^{-1}_\ell A_{\rm{irr},\ell}V_{\rm{uc}}]-2\mathbf{r}_{\rm{uc}}^TA_{\rm{irr},\ell}^TD_\ell^{-1}A_{\rm{irr},\ell}\mathbf{r}_{\rm{uc}}\right) \\
    &\Pi_{\rm{uc}}=2{\rm{Tr}}[A_{\rm{irr}}]+\sum_\ell \left(2{\rm{Tr}}[A_{\rm{irr},\ell}^TD^{-1}_\ell A_{\rm{irr},\ell}V_{\rm{uc}}]+2\mathbf{r}_{\rm{uc}}^TA_{\rm{irr},\ell}^TD^{-1}_\ell A_{\rm{irr},\ell}\mathbf{r}_{\rm{uc}}\right)+\frac{1}{2}\rm{Tr}[V_{\rm{uc}}^{-1}D],
\end{align}
where $\mathbf{r}_{\rm{uc}}$ are the first cumulants. It is important to note that the separation of the different terms stemming from the different output channels is crucial to correctly assess the thermodynamics of the system and to not overestimate the entropy production (cf.~\cite{PhysRevE.82.011143_SI}). 

For the conditional evolution, i.e., when the measurement outcomes are recorded, the entropy production and flux rates become stochastic variables on the single trajectory level. Their expressions are given in Ref.~\cite{belenchia2019entropy_SI} and read:
\begin{align}
    & \phi_{c,\mathbf{r}}=-{\rm{Tr}}[A_{\rm{irr}}]dt-\sum_\ell \left(2{\rm{Tr}}[A_{\rm{irr},\ell}^TD^{-1}_\ell A_{\rm{irr},\ell}V]dt+2\mathbf{r}^TA_{\rm{irr},\ell}^TD_\ell^{-1}A_{\rm{irr},\ell}\mathbf{r}dt\right)\\
    & \pi_{c,\mathbf{r}}=2{\rm{Tr}}[A_{\rm{irr}}]dt+\sum_\ell \left(2{\rm{Tr}}[A_{\rm{irr},\ell}^TD_\ell^{-1}A_{\rm{irr},\ell}V]dt+2\mathbf{r}^TA_{\rm{irr},\ell}^TD_\ell^{-1}A_{\rm{irr},\ell}\mathbf{r}dt\right)+\frac{1}{2}\rm{Tr}[V^{-1}(D-\chi)],
\end{align}
where $\bm{r}$ is the conditional first moment. 
All the stochasticity in the above expressions is encoded in $\bm{r}$.

Averaging over all the trajectories in phase-space we obtain deterministic entropy production and flux rates~\cite{belenchia2019entropy_SI}
\begin{IEEEeqnarray}{rCl}
     \Phi_{\rm{c}} 
     &=&-{\rm{Tr}}[A_{\rm{irr}}]-\sum_\ell\left(2{\rm{Tr}}[A_{\rm{irr},\ell}^TD^{-1}_\ell A_{\rm{irr},\ell}V_{\rm{uc}}]+2\mathbf{r}_{\rm{uc}}^TA_{\rm{irr},\ell}^TD_\ell^{-1}A_{\rm{irr},\ell}\mathbf{r}_{\rm{uc}}\right) \\[0.2cm]
    &=& \Phi_{\rm{uc}}, \nonumber\\[0.2cm]
    \Pi_{\rm{c}}
    &=& 2{\rm{Tr}}[A_{\rm{irr}}]+\frac{1}{2}\rm{Tr}[V_{\rm{uc}}^{-1}D]+\frac{1}{2}{\rm{Tr}}[(V^{-1}-V_{\rm{uc}}^{-1})D-V^{-1}\chi]
    +\sum_\ell \left(2{\rm{Tr}}[A_{\rm{irr},\ell}^TD^{-1}_\ell A_{\rm{irr},\ell}V_{\rm{uc}}]+2\mathbf{r}_{\rm{uc}}^TA_{\rm{irr},\ell}^TD^{-1}_\ell A_{\rm{irr},\ell}\mathbf{r}_{\rm{uc}}\right) \nonumber\\[0.2cm]
    &=&\Pi_{\rm{uc}}+\dot{\mathcal{I}}, 
\end{IEEEeqnarray}
where 
\begin{equation}\label{SM_Idot}
    \dot{\mathcal{I}}=\frac{1}{2}\rm{Tr}[V^{-1}(D-\chi(V))-V_{\rm{uc}}^{-1}D].
\end{equation} 
As can be seen, the conditional entropy flux, on average, coincides with the unconditional one. 
For the entropy production, on the other hand, this is not the case.
Instead, the mismatch between the two is given by the net information gain $\dot{\mathcal{I}}$.

\section{\label{sec:level2} Optomechanical cavity system}
In this section, we apply the above ideas to the specific experiment reported in the main text. In particular, we recognise the steady-state of the unconditional dynamics as a non-equilibrium steady-state (NESS) and derive the expression for the unconditional and conditional entropy production and flux rates. 

The experimental system considered in the main text is a cavity optomechanical system working in a range of parameters in which the adiabatic elimination of the cavity field gives a good description of the dynamics of the mechanical system, i.e., in the bad-cavity weak coupling limit $\kappa\gg\Omega_m\gg g$, where $\kappa$ is the cavity linewidth, $\Omega_m$ is the mechanical resonance frequency and $g$ the linearised, multi-photon optomechanical coupling. 

\subsection{\label{sec:level2} Master equation in adiabatic approximation and non-equilibrium steady-state}
As detailed in~\cite{hofer2017quantum_SI}, by adiabatically eliminating the cavity mode the dynamics of the system can be enormously simplified leading to a master equation of the form
\begin{equation}\label{adlimit}
    \dot{\rho}_m=-i\left[\delta\Omega_m c^\dag c,\rho_m\right]+\left(\Gamma_m (\bar{n}_{\rm{th}}+1)+\Gamma_-\right)\mathcal{D}[c]\rho_m+\left(\Gamma_m \bar{n}_{\rm{th}}+\Gamma_+\right)\mathcal{D}[c^\dag]\rho_m,
\end{equation}
where $\rho_m$ represents the state of the sole mechanical mode and we have introduced the following quantities
\begin{align}
    &\delta\Omega_m=g^2 \rm{Im}[\eta_- +\eta_+]\\
    &\eta_\pm=\frac{1}{\kappa/2+i(-\Delta\pm\Omega_m)}\\
    &\Gamma_\pm=2g^2\rm{Re}[\eta_\pm],
\end{align}
with $\Delta$ the effective laser detuning. The master equation of the system studied in the main text is obtained in the limit in which the detuning vanishes and, introducing also the stochastic terms due to the continuous monitoring, it is exactly of the form Eq.\eqref{SMEgen}. We report it here for later convenience  
\begin{equation}\label{SMEad}
    d\rho_c=\mathcal{L}_{\rm{th}}\rho_c dt +\Gamma_{\rm{qba}} \left(\mathcal{D}(c)+\mathcal{D}(c^\dag)\right)\rho_c dt+\sqrt{\eta\Gamma_{\rm{qba}}}\left(\mathcal{H}(X)\rho_c dW_X+\mathcal{H}(Y)\rho_c dW_Y\right),
\end{equation}
where $\mathcal{L}_{\rm th}\rho=\Gamma_m (\bar{n}_{\rm{th}}+1)\mathcal{D}[c]\rho+\Gamma_m \bar{n}_{\rm{th}}\mathcal{D}[c^\dag]\rho$ represents the contact with the mechanical phonon thermal bath,  which is not subject to continuous monitoring. This equation was first introduced in~\cite{PhysRevLett.107.213603_SI} and then formally derived in~\cite{doherty2012quantum_SI}.

Given the master equation, and the initial condition of the dynamics described in the main text, the first and second cumulants of the quadratures of the mechanical oscillator evolve according to 
\begin{eqnarray}\label{e:sme_moments}
d\rfw(t) &=& -\frac{\Gm}{2}\rfw(t) dt + \sqrt{4\eta_{\rm{det}}\Gamma_{\rm{qba}}} V(t) \dW,\\
\dot{V}(t) &=& -\Gm V(t) + \Gm (\bar{n}_{\rm{th}}+1/2)+\Gamma_{\rm{qba}} - 4\eta_{\rm{det}}\Gamma_{\rm{qba}} V(t)^2,\label{e:V_c_dynamics}
\end{eqnarray}
where $\dW=(dW_x,dW_y)^T$ and the covariance matrix is always diagonal $\mathbf{V}(t)$=V(t)\openone.  

Let us focus for a moment on the unconditional dynamics. In the experimental realisation discussed in the main text, the initial state of the mechanical oscillator corresponds to the steady-state of the unconditional dynamics, characterised by $\mathbf{r}_\mathrm{uc} = \mathbf{0}$ and $\Vuc =( \bnth + 1/2 + \Gqba/\Gm)\1$. It should be noted here that, in  the unconditional steady-state the system has thermalised at an effective temperature different from the one of the phononic bath. Crucially, this is not an equilibrium state due to the fact that the mechanical phonon bath ($\mathcal{L}_{\rm{th}}$) and the optical mode ($\mathcal{L}_{\rm{qba}}$) represent two different baths giving rise to a steady energy current. Indeed, consider the energy fluxes to each bath at the unconditional steady-state. These can be obtained by looking at
\begin{align}
\frac{d}{dt}\langle  c^\dag c\rangle&=\rm{Tr}[ \mathcal{L}_{m}c^\dag c]+\rm{Tr}[ \mathcal{L}_{\rm{qba}}c^\dag c]=\Gamma_m (\langle  c^\dag c\rangle-\bar{n}_{\rm{th}})+\Gamma_{\rm{qba}}.
\end{align}
While it is easy to show that at the unconditional steady-state the two energy fluxes cancel each other, the fact that they are not individually zero tells us that there is a steady-current making all the energy coming from the thermal phonon bath flow into the optical mode, i.e., the unconditional steady-state is a non-equilibrium steady-state (NESS). This implies that for the unconditional dynamics $\Phi_{\rm{uc}}=-\Pi_{\rm{uc}}\neq 0,\,\,\forall t$. 

\subsection{\label{sec:level2} Irreversible entropy production}
In order to compute the non-zero unconditional entropy rates, we need to resort to Eq.\eqref{ucthermo} and consider the contributions of the three output channels related to the phonon bath and the two independent Wigner increments ($dW_X,\,dW_Y$)  in  the master equation. 
This is justified from the fact that, while the phonon thermal bath is not measured, the optical bath is probed via two distinct output channels characterised by independent Wiener increments. However, care has to used when considering the master equation in~\eqref{SMEad} in conjuction with Eq.\eqref{ucthermo}. Indeed, it is easy to realise that a n\"aive use of such equations leads to a non-zero entropy flux to the thermal phonon bath but to a vanishing one to the optical bath. However, we have already seen that there is a non-vanishing {\it energy} flux associated with the optical bath, to which there must  correspond a non-vanishing entropy flux and an associated entropy production term.
This discrepancy can be traced back to the singular character of some of the matrices entering Eq.\eqref{ucthermo} and is linked to the fact that the adiabatic elimination of the cavity field entails a loss of information on the details of the process undergone by the system-bath compound. While this is not critical when evaluating properties of the state of the mechanical system, the evaluation of thermodynamic quantities — which are instead crucially dependent on the process itself — should be done {\it cum grano salis}.

There are two alternative ways to recover a consistent result. On the one hand, it is possible to work at the level of Eq.~\eqref{SMEad} but consider the full system-plus-environment interaction and evolution in a microscopic collisional model picture. The collisional model allows to derive, from a microscopic perspective, a non-null entropy flux with the optical bath even after the adiabatic elimination and the RWA have been performed. This points, again, toward the fact that the knowledge of the sole master equation, while sufficient to completely characterise the dynamics of the system, does not capture entirely the thermodynamics of the problem~\cite{De_Chiara_2018_SI} in general. A full derivation of the fluxes from a microscopic collisional model is beyond the scope of this work, and will be discussed in detail in an upcoming work~\cite{CMarticle_SI}.

On the other hand, the expressions for the needed entropy flux and production rates can be derived also by considering the full composite system of cavity+mechanical mode before the adiabatic elimination of the cavity mode and the RWA. In this way, it is possible to correctly obtain the entropy flux associated to the optical bath --- responsible for the term $\Gamma_{qba}(\mathcal{D}[c]+\mathcal{D}[c^\dag])$ in the master equation after the adiabatic elimination --- which is, indeed, not vanishing even in the bad-cavity limit.  

\subsubsection{Full optomechanical system}
The master equation describing the unconditional dynamics of the composite mechanical-optical modes system (we refer to~\cite{hofer2017quantum_SI} for further details) is given by
\begin{equation}\label{fullME}
    \dot{\rho}=-i\left[H_{\rm{lin}},\rho\right]+\kappa\mathcal{D}[a]\rho+\Gamma_m (\bar{n}+1)\mathcal{D}[c]\rho+\Gamma_m \bar{n}\mathcal{D}[c^\dag]\rho,
\end{equation}
where $\rho$ represents now the state of the composite system, $H_{\rm{lin}}=\Omega_m c^\dag c+g(a+a^\dag)(c+c^\dag)-\Delta a^\dag a$ is the linearised optomechanical Hamiltonian with detuning $\Delta$, multi-photon optomechanical coupling $g$ and the operators $\hat{a},\hat{a}^\dag$ are the annihilation and creation operators for the cavity optical mode. Moreover, in line with the notation in the main text, $\kappa$ is the linewidth of the optical cavity, $\Gamma_m$ is the mechanical damping factor, and $\bar{n}_{\rm{th}}$ is the mean number of excitations in the thermal phononic bath in contact with the mechanical mode. The output mode of the cavity is measured with a balanced homodyne detection scheme in order to gather information on the position of the mechanical mode. This corresponds to the stochastic master equation
\begin{equation}\label{fullSME}
    d{\rho}=-i\left[H_{\rm{lin}},\rho\right]dt+\kappa\mathcal{D}[a]\rho dt+\Gamma_m (\bar{n}_{\rm{th}}+1)\mathcal{D}[c]\rho dt+\Gamma_m \bar{n}_{\rm{th}}\mathcal{D}[c^\dag]\rho dt+\sqrt{\eta_{\rm{det}} \kappa}\mathcal{H}[-i a]\rho dW,
\end{equation}
where $\mathcal{H}[\hat{A}]\rho=A\rho+\rho A^\dag -\rm{Tr}[A\rho+\rho A^\dag]\rho$, $\eta_{\rm{det}}$ is the efficiency of the detector, and $dW$ is a real Wiener increment. 

The equivalent description of the system in terms of first and second cumulants of the Wigner distribution in phase-space is obtained by using the following matrices [(c.f. Eqs.~\eqref{r_riccatti}-~\eqref{V_riccatti}]:
\begin{align}
A_{H_{\rm{lin}}}&=\begin{pmatrix}
    0 & \Omega_m & 0 & 0\\
    -\Omega_m & 0 & -2g & 0\\
    0 & 0 & 0 & -\Delta\\
    -2g & 0 & \Delta & 0
    \end{pmatrix} \label{matrices1} \\ 
    A_{\rm{irr,th}}&=\begin{pmatrix}
    -\Gamma_m/2 & 0 & 0 & 0\\
    0 & -\Gamma_m/2 & 0 & 0\\
    0 & 0 & 0 & 0\\
    0 & 0 & 0 & 0
    \end{pmatrix} && A_{\rm{irr,opt}}=\begin{pmatrix}
    0 & 0 & 0 & 0\\
    0 & 0 & 0 & 0\\
    0 & 0 & -\kappa/2 & 0\\
    0 & 0 & 0 & -\kappa/2
    \end{pmatrix}  \label{matrices2} \\ 
 D_{\rm{th}}&=\begin{pmatrix}
    \Gamma_m(\bar{n}_{\rm{th}}+1/2) & 0 & 0 & 0\\
    0 & \Gamma_m(\bar{n}_{\rm{th}}+1/2) & 0 & 0\\
    0 & 0 & 0 & 0\\
    0 & 0 & 0 & 0
    \end{pmatrix} && D_{\rm{opt}}=\begin{pmatrix}
    0 & 0 & 0 & 0\\
    0 & 0 & 0 & 0\\
    0 & 0 & \kappa/2 & 0\\
    0 & 0 & 0 & \kappa/2
    \end{pmatrix};\\
    \label{matrices3} \\ 
 C&=\begin{pmatrix}
    0 & 0 & 0 & 0\\
    0 & 0 & 0 & 0\\
    0 & 0 & 0 & 0\\
    0 & 0 & 0 & \sqrt{2\kappa\eta_{\rm{det}}}
    \end{pmatrix} && \Gamma=\begin{pmatrix}
    0 & 0 & 0 & 0\\
    0 & 0 & 0 & 0\\
    0 & 0 & 0 & 0\\
    0 & 0 & 0 & -\sqrt{\kappa\eta_{\rm{det}}/2}
    \end{pmatrix};\label{matrices4}
\end{align}
Given these expressions, in conjunction with Eqs.\eqref{ucthermo}, the expressions for the unconditional entropy flux and production rate in the bad-cavity limit, i.e. for the system of interest as described by Eq.\eqref{SMEad}, can be easily determined. For the entropy fluxes of the $X$ and $Y$ channels we have
\begin{equation}
 \Phi_{\rm{uc},X}=\Phi_{\rm{uc},Y}=-2 V_{\rm{uc}} \Gamma_{\rm{qba}},
\end{equation}
where we have used the fact that, in the case of interest, the unconditional first momenta vanishes and that the unconditional steady-state of the full dynamics has a covariance matrix proportional to the identity $V_{\rm{uc},11}=V_{\rm{uc},22}\equiv V_{\rm{uc}}$ and $V_{\rm{uc},12}=0$. The corresponding contribution to the unconditional entropy production rate are instead given by
\begin{equation}
 \Pi_{\rm{uc},X}=\Pi_{\rm{uc},Y}= \Gamma_{\rm{qba}}\left(2 V_{\rm{uc}}+\frac{1}{2V_{\rm{uc}}}\right).
\end{equation}
Summing up these rates with the ones due to the thermal phonon bath, the  final result for the case of interest in the main text is 
\begin{align}
 \Phi_{\rm{uc}}&=\Phi_{\rm{uc, th}}+\Phi_{\rm{uc},X}+\Phi_{\rm{uc},Y}\\
 &=\Gamma_m-\frac{V_{\rm{uc}}}{\bar{n}_{\rm{th}}+1/2}\Gamma_{m}-4 V_{\rm{uc}} \Gamma_{\rm{qba}},
\end{align}
and 
\begin{align}
 \Pi_{\rm{uc}}&=\Pi_{\rm{uc, th}}+\Pi_{\rm{uc},X}+\Pi_{\rm{uc},Y}\\
 &=-2\Gamma_m+\frac{V_{\rm{uc}}}{\bar{n}_{\rm{th}}+1/2}\Gamma_{m}+\frac{(\bar{n}_{\rm{th}}+1/2)\Gamma_m}{V_{\rm{uc}}}+4 V_{\rm{uc}} \Gamma_{\rm{qba}}+\frac{\Gamma_{\rm{qba}}}{V_{\rm{uc}}},
\end{align}
As a sanity check,  using the form of the unconditional steady-state we find $\Phi_{\rm{uc}}=-\Pi_{\rm{uc}}$ as indeed expected.

\subsubsection{Information gain}
For what concerns the conditional dynamics, we have already seen that $\Pi_{\rm{c}}=\Pi_{\rm{uc}}+\dot{\mathcal{I}}$ and $\Phi_{\rm{c}}=\Phi_{\rm{uc}}$, so that we only need to derive the net information gain $\dot{\mathcal{I}}$ in Eq.\eqref{SM_Idot}. Considering that the covariance matrix of the conditioned dynamics remains proportional to the identity at all times, we find 
\begin{align}
\dot{\mathcal{I}}&=\frac{1}{2}\rm{Tr}[V^{-1}(D-\chi(V))-V_{\rm{uc}}^{-1}D]\\
&=\frac{\Gm \Vuc - 4\eta_{\rm{det}}\Gamma_{\rm{qba}} V(t)^2}{V(t)}-\Gm
\end{align}
Finally, let us stress once again that $\dot{\mathcal{I}}$ is minus the rate at which information is acquired by performing the measurement and tracking the outcomes~\cite{belenchia2019entropy_SI}. This quantity is, by definition, zero at the steady-state of the conditional evolution. This is intuitive since, at the conditional steady-state, the continuous measurement is not adding any more information with respect to what it was acquired until that point. 

However, it is crucial to note that the measurement process is still necessary to maintain the conditional steady-state. Turning off the detector's efficiency would drive the system back to the unconditional steady-state. This fact is encoded by the innovation matrix $\chi(V)$ and the corresponding term 
\begin{equation}
    \mathcal{G}(t) := -\rm{Tr}[V^{-1}\chi(V)]/2,
\end{equation} 
in $\dot{\mathcal{I}}$ [Eq.~\eqref{SM_Idot}]. 
This term can be interpreted as the information gained by acquiring the measurement result at time $t$ with respect to the knowledge given by the outcome record up to time $t-dt$. This is what is called \textit{differential} gain of information in the main text. This quantity does not vanish at steady-state. It represents the on-going effort of the continuous measurement to keep the system in the conditional steady-state~\cite{CMarticle_SI}. 



\section{Estimation of the dynamical covariance matrix}
Here we calculate the variance of the difference trajectory, $\mathbf{d}(t) = \rfw(t)-\rbw(t)$, as done in Ref.~\cite{PhysRevLett.123.163601_SI}, to the case of time-dependent predicted trajectory $\rfw(t)$.

Experimentally, we can extract from the measurement the predicted and retrodicted trajectories, $\rfw(t)$ and $\rbw(t)$ respectively~\cite{PhysRevLett.123.163601_SI}. The variance of the ensemble of difference trajectory $\mathbf{d}(t)$ is
\begin{eqnarray}\label{e:V_D_formal}
\Vd(t) \equiv \mathbb{E}\left[(\rfw(t) - \rbw(t))^2\right]  = \mathbb{E}\left[\rfw(t)^2\right] + \mathbb{E}\left[\rbw(t)^2\right] - 2\mathbb{E}\left[\rfw(t)\rbw(t)\right].
\end{eqnarray}
The formal solutions for the forward and backward trajectories are
\begin{subequations}
\begin{align}
    \rfw(t) &= \sqrt{4\Gmeas}\int_0^t V(s) e^{-\frac{\Gm}{2}(t-s)}\dW(s),\\
    \rbw(r) &= 4\Gmeas V_E \int_t^\infty e^{\lambda(t-s)}\rfw(s) ds + \sqrt{4\Gmeas} V_E \int_t^\infty e^{\lambda(t-s)}\dW (s),
\end{align}
\end{subequations}
where $\Gmeas= \eta_{\rm{det}} \Gamma_\text{qba}$, $\lambda = 4\Gmeas V_E - \Gm/2$ and $V_E = V + \Gm/(4\Gmeas)$. For the forward trajectory, the initial conditions are $\rfw(0) = 0$ and $V(0)=\Vuc$.
For the backward trajectory, we assume that the initial conditions are far in the future, such that at time $t$ they are forgotten. This also implies that the backward conditional variance $V_E$, at time $t$, has reached the steady state.

We move now to calculating the three terms in Eq.~\eqref{e:V_D_formal}, considering the general case where the conditional variance $V(t)$ is not at the steady state. This generalises the case considered in~\cite{PhysRevLett.123.163601_SI}.
For the first term, we start by calculating the two-time correlator for the forward trajectory,  
\begin{eqnarray}\label{e:corr_rfw}
\mathbb{E}\left[\rfw(t)\rfw(\tau)\right]  = 4\Gmeas e^{-\frac{\Gm}{2}(t+\tau)} \int_0^{\mathrm{min}(t,\tau)} e^{\Gm s} V(s)^2 ds.
\end{eqnarray}
We can compute the first term by taking $t=\tau$, then
\begin{eqnarray}\label{e:var_rfw}
\mathbb{E}\left[\rfw(t)^2\right]  = 4\Gmeas e^{-\Gm t} \int_0^t e^{\Gm s} V(s)^2 ds.
\end{eqnarray}
The second term has been already calculated in~\cite{PhysRevLett.123.163601_SI}, and is given by
\begin{eqnarray}\label{e:var_rbw}
\mathbb{E}\left[\rbw(t)^2\right] = 4\Gmeas V_E^2/\Gm = V_E + \Vuc.
\end{eqnarray}
Finally, for the third term, we have
\begin{eqnarray}\label{e:cross_corr_rfw_rbw}
 \mathbb{E}\left[\rfw(t)\rbw(t)\right] = 4\Gmeas V_E \int_t^{\infty} e^{\lambda (t-s)} \mathbb{E}\left[\rfw(s)\rfw(t)\right] ds = 4\Gmeas e^{-\Gm t} \int_0^t e^{\Gm s} V(s)^2 ds.,
\end{eqnarray}
where we have used Eq.~\eqref{e:corr_rfw} and the property that Wiener increments at different time intervals are uncorrelated.

The integral in Eqs.~\eqref{e:var_rfw} and \eqref{e:cross_corr_rfw_rbw} can be simplified by formally integrating Eq.~\eqref{e:V_c_dynamics}, from which
\begin{eqnarray}\label{e:int_from_V_c_dynamics}
4\Gmeas\int_0^t e^{\Gm (s-t)} V(s)^2 ds = \Vuc - V(t).
\end{eqnarray}
Combining Eqs.~\eqref{e:V_D_formal}, \eqref{e:var_rbw}, \eqref{e:var_rfw}, \eqref{e:cross_corr_rfw_rbw} and \eqref{e:int_from_V_c_dynamics} together we obtain
\begin{eqnarray}\label{e:Vd_final}
\Vd(t) = V(t) + V_E = V(t) + V + \Gm/(4\Gmeas).
\end{eqnarray}

The conditional variance $V(t)$ can be estimated, then, from the experimentally accessible variance $\Vd(t)$, after the small offset value $V + \Gm/(4\Gmeas)$ is subtracted. In particular, in the large cooperativity ($\Gqba>>\Gm$) and large detection efficiency ($\eta_{\rm{det}}\approx 1$) limits, relevant to the experiment described in the main text, the offset $V + \Gm/(4\Gmeas)\approx V$, where $V$ is the steady state value of the conditional variance $V(t)$. Then, from Eq.~\eqref{e:Vd_final}, we have $V(t) \approx \Vd(t) - V$, where $V$ can be estimated from the steady state of the experimental $\Vd(t)$ using the fact that $\Vd(\infty)\approx 2V$ (see~\cite{PhysRevLett.123.163601_SI} for the derivation of this result).

\end{document}